\documentclass[letterpaper,12pt]{article}   
\usepackage{osajnl2} 
\usepackage[draft]{hyperref} 
\usepackage{amsmath, amssymb, amsthm}
\usepackage{setspace}

\begin{document}

\title{Tunable Hybridization at Mid Zone and Anomalous Bloch-Zener Oscillations in Optical Waveguide Ladders}

\author{Ming Jie Zheng,$^{1}$ Gang Wang,$^{1,2,*}$ and Kin Wah Yu$^{1,3,**}$}

\address{$^1$Department of Physics, The Chinese University of Hong Kong, Shatin, New Territories, Hong Kong, China}
\address{$^2$Department of Physics, Fudan University, Shanghai 200433, China}
\address{$^3$Institute of Theoretical Physics, The Chinese University of Hong Kong, Shatin, New Territories, Hong Kong, China}
\address{$^{*}$Present Address: College of Science, China University of Mining and Technology, Xuzhou 221008, China}
\address{$^{**}$E-mail: kwyu@phy.cuhk.edu.hk}

\begin{abstract}

We have studied the optical oscillation and tunneling of light waves in optical waveguide ladders formed by two coupled planar optical waveguide arrays. For the band structure, a mid-zone gap is formed due to band hybridization and its wavenumber position can be tuned throughout the whole Brillouin zone, which is different from the Bragg gap. By imposing a gradient in the propagation constant in each array, Bloch-Zener oscillation (BZO) is realized with Zener tunneling between the bands occurring at mid zone, which is contrary to the common BZO with tunneling at the center or edge of the Brillouin zone. The occurrence of BZO is demonstrated by using the field-evolution analysis. The tunable hybridization at mid zone enhances the tunability of BZO in the optical waveguide ladders. This work is of general and fundamental importance in understanding the coherent phenomena in lattice structures.

\end{abstract}

\date{\today}


\maketitle

\newpage


Light propagation in optical waveguide arrays (OWA) is important both in fundamental science and practical applications \cite{Nature.424.817.2003}. Optical waves propagating in a discrete photonic structure behave analogously to electrons moving in a semiconductor crystal \cite{Nature.424.817.2003}.
Bloch oscillation (BO) \cite{OptLett.23.1701.1998, PhysRevLett.83.4752.1999, PhysRevLett.83.4756.1999} and Zener tunneling (ZT) \cite{PhysRevLett.96.023901.2006, PhysRevLett.96.053903.2006} are two fundamental transport phenomena in current research \cite{PhysRep.366.103.2002, PhysRevLett.102.076802.2009}. BO is the oscillatory motion of particles in a periodic potential under an external force \cite{ZPhys.52.555.1928}, whose optical equivalent is a linear gradient in refractive index \cite{PhysRevLett.83.4756.1999, PhysRevLett.102.076802.2009, PhysRevLett.91.263902.2003, PhysRevA.81.033829} or geometrical variation in waveguides \cite{OptLett.31.1651.2006}. In multiband models, ZT acts as the regular outbursts of radiation from BO into higher-order bands \cite{RSocLondA.145.523.1934, PhysRevLett.96.023901.2006}. The superposition of BO and ZT causes a double-periodic motion of wavepackets, which is referred as the Bloch-Zener oscillation (BZO) \cite{NewJPhys.8.110.2006, News.Longhi}. BZO has been predicted theoretically for classical \cite{EurophysLett.76.416.2006} and nonclassical light \cite{PhysRevLett.101.193902.2008} in waveguide arrays and demonstrated experimentally in binary superlattices \cite{PhysRevLett.102.076802.2009}. In most of previous works on BZO, ZT usually occurs at the center or edge of the Brillouin zone, that is, tunneling only takes place at specific values of wavenumber. This condition restricts the tunability of BZO in applications. To improve the tunability, we analyze the underlying physical mechanism for the light propagation starting from the dispersion relation, because the field evolution is completely controlled by the dispersion relation \cite{PhysRep.463.1.2008}.

\begin{figure}[htbp]
  \centering
  \includegraphics[width=0.8 \textwidth]{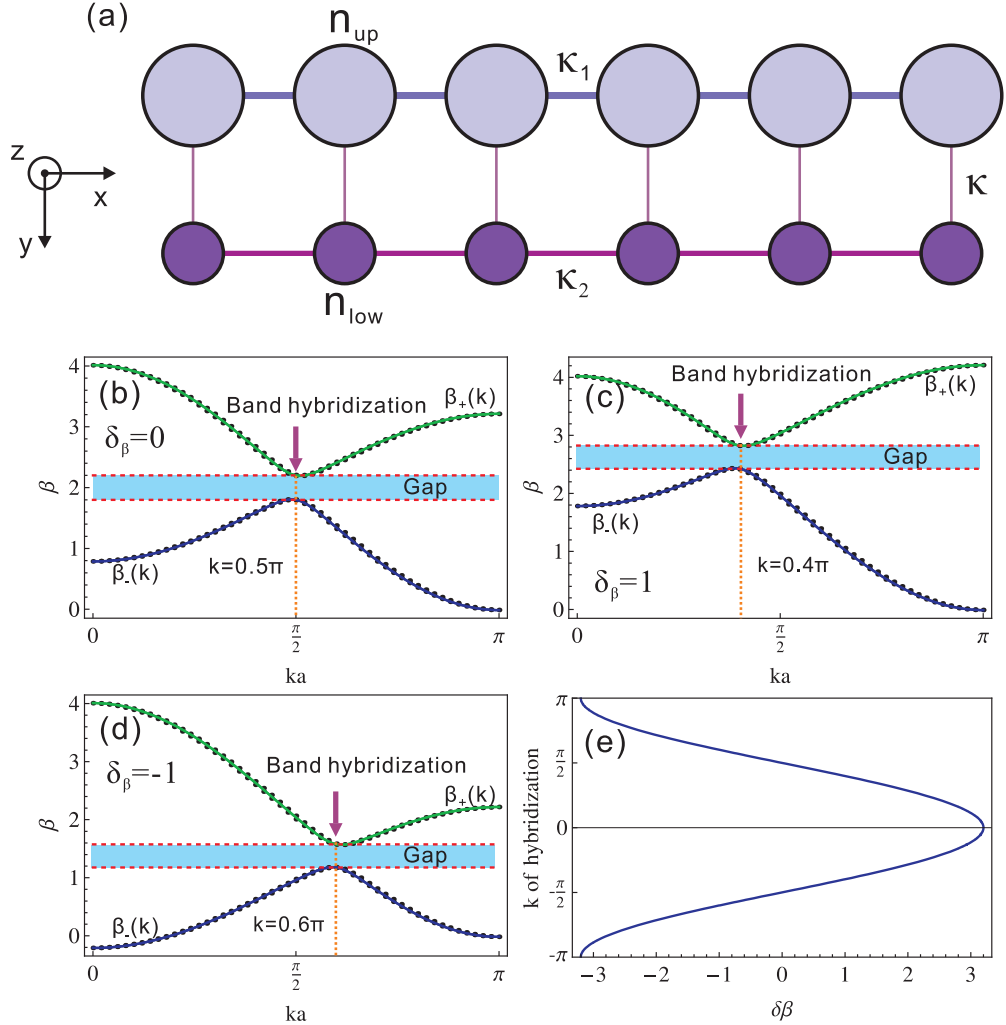}
  \caption{(Color online) (a) Schematic diagram of a cross section of an optical waveguide ladder. The coupling constants for the positive/negative coupling case are $\kappa_1 = 1$, $\kappa_2 = -0.5$, and $\kappa = 0.2$. Band structures for various propagation constant differences (b) $\delta_\beta = 0$, (c) $\delta_\beta = 1$, and (d) $\delta_\beta = -1$. Band hybridization occurs at $k = 0.5\pi$, $0.4\pi$, and $0.6\pi$, respectively. The analytic results (solid lines) match well with the numerical results (discrete dots). (e) Wavenumber position of band hybridization as a function of propagation constant differences $\delta_\beta$ between two arrays.}
  \label{fig:OWLband}
  \end{figure}

In this Letter, we consider an optical waveguide ladder (OWL) formed by two coupled planar optical waveguide arrays as shown schematically in Fig.~\ref{fig:OWLband}. The coupling constants between neighboring waveguides in the upper and lower arrays are $\kappa_1$ and $\kappa_2$, respectively. The inter-array coupling constant is $\kappa$. In this work, both $\kappa_1$ and $\kappa$ are positive. There are two cases for $\kappa_2$: (i) $\kappa_2 < 0$ for the positive/negative coupling (PNC) case and (ii) $\kappa_2 > 0$ for the positive/positive coupling (PPC) case. Although the positive and negative coupling constants can be realized by using dielectric and metal OWAs \cite{OptLett.30.2894.2005, OptLett.31.1322.2006, PhysRevLett.97.073901.2006}, respectively, metal OWAs are incompatible with dielectric OWAs due to losses. the PNC is much harder to realize experimentally than the PPC, which can be obtained by pure dielectric OWAs. However, there is no ``true band gap'' in the PPC case, while a ``true band gap'' indeed exists in the PNC case. Therefore, we will use the PNC case to illustrate the dynamics of BZO, and compare its results with those in PPC case. The numbers of waveguides in the upper and lower arrays are the same $N_{\rm up} = N_{\rm low} = 100$, and the separation between two nearest-neighbor waveguides is assumed to be $1$. To ensure the occurrence of BO and ZT, we impose a linear gradient in the propagation constant, which can be realized by either a temperature gradient in thermo-optical waveguide arrays \cite{PhysRevLett.83.4752.1999, PhysRevLett.96.023901.2006} or a gradient in effective refractive index by electro-optical effects \cite{OptLett.23.1701.1998, PhysRevA.81.033829}. Light propagates along the waveguide ($z$) axis.


The propagation of light in the OWL is described by a set of coupled differential equations of modal amplitudes $a_n$ ($b_n$) in the upper (lower) array,
\begin{equation}\label{eq:EqMotion}
\begin{aligned}
(\mathbf{i}\frac{d}{dz} + \beta_{u})a_n(z)+ \kappa_1\left[a_{n+1}(z)+a_{n-1}(z)\right] + \kappa b_n(z) & =0\,, \\
(\mathbf{i}\frac{d}{dz} + \beta_{l})b_n(z)+ \kappa_2\left[b_{n+1}(z)+b_{n-1}(z)\right] + \kappa a_n(z) & =0\,,
\end{aligned}
\end{equation}
where $\beta_{u} = \beta_0 + \alpha_{u} n$ and $\beta_{l} = \beta_0 + \delta_\beta + \alpha_{l} n$ are the onsite propagation constants of the $n$th waveguide with gradient $\alpha_{u}$ and $\alpha_{l}$ in the upper and lower arrays, respectively. We apply the same gradient in the two arrays $\alpha_u = \alpha_l$. The propagation constant difference between the two arrays is $\delta_\beta$. The two $\kappa$ terms in Eq.~(\ref{eq:EqMotion}) represent the inter-array couplings between the two waveguide arrays. Substituting the solutions $a_n^m(z)=u_n^m e^{\mathbf{i}\beta_m z}$ and $b_n^m(z)=v_n^m e^{\mathbf{i}\beta_m z}$ into Eq.~(\ref{eq:EqMotion}), we obtain a matrix equation $\mathbf{H} |m\rangle = \beta |m\rangle$, where the Hamiltonian matrix is defined as $\mathbf{H} = \{ \{\mathbf{U}, \mathbf{Z} \}, \{\mathbf{Z}, \mathbf{V} \} \}$. The matrix elements are $U_{i,i} = \beta_u$, $U_{i,i-1} = U_{i,i+1}=\kappa_1$, $V_{i,i} = \beta_l$, $V_{i,i-1} = V_{i,i+1}=\kappa_2$, $Z_{i,i} =\kappa$, and other elements are zero. The column vectors $|m\rangle$ and $\beta$ are eigenvectors and eigenvalues of $\mathbf{H}$, respectively.
In the absence of gradient in both arrays, $\alpha_{u} = \alpha_{l} = 0$, we can obtain a dispersion relation
\begin{equation}\label{eq:DISP}
\beta_{\pm}=\beta_0+\frac{\delta_\beta}{2}+(\kappa_1 +\kappa_2)\cos k \pm \left\{\kappa^2+\left[(\kappa_1-\kappa_2)\cos k -\frac{\delta_\beta}{2}\right]^2\right\}^{\frac{1}{2}}\,.
\end{equation}

The dispersion relation Eq.~(\ref{eq:DISP}) relates the longitudinal wavenumber $\beta$ to the transverse wavenumber $k$ and determines the evolution of the normal modes during propagation \cite{PhysRep.463.1.2008}. The band structures for PNC case obtained from Eq.~(\ref{eq:DISP}) are shown in Figs.~\ref{fig:OWLband}(b)-\ref{fig:OWLband}(d) for various values of propagation constant differences $\delta_\beta$ between two arrays. The analytic results (solid lines) match with the numerical results (dots). There are two minibands $\beta_{-}$ and $\beta_{+}$ with a band gap opening at the mid zone ($0 \leq k \leq \pi$). The physical origin of the mid-zone gap is due to the hybridization of the upper array band and the lower array band. The total propagation constant difference between the two bands is defined as $\Delta  = |\beta_{+} - \beta_{-}|$. The band hybridization occurs when the propagation constant difference reaches its minimum, that is, $d\Delta/dk = 0$. The wavenumber $k$ depends on $\kappa$ and $\delta_{\beta}$. If there is no coupling between two arrays ($ \kappa = 0$), no band gap will be opened. The nonzero inter-array couplings ($\kappa \neq 0$) through evanescent fields between the two arrays can cause the two bands to split at the degenerate point due to level anticrossing, which leads to two separated minibands with a mid-zone gap. The wavenumber position of the band gap can be tuned by varying the propagation constant differences $\delta_{\beta}$ between two arrays. As shown in Figs.~\ref{fig:OWLband}(b)-\ref{fig:OWLband}(d), the band hybridization occurs at $k = 0.5 \pi$, $0.4\pi$, and $0.6\pi$ when $\delta_{\beta} = 0$, $1$, and $-1$, respectively. In Fig.~\ref{fig:OWLband}(e), the wavenumber position of the band hybridization covers the whole range of $[-\pi,\pi]$ when $\delta_{\beta}$ varies from $-3$ to $3$. For a certain $\delta_{\beta}$, the wavenumber position of the band hybridization is symmetric with respect to $k=0$. The gap due to tunable band hybridization is different from the usual band gap appearing at the center or edge of the Brillouin zone due to Bragg reflection. The latter is tunable vertically in the gap width, as in binary superlattices \cite{PhysRevLett.102.076802.2009}. While the former can be tuned not only in the gap width but also in the wavenumber position.

In OWL with the obtained band structure, the gap due to band hybridization is so small that Zener tunneling between the two minibands is possible. If we impose a gradient in the propagation constants in each array, BO and breathing-wave-like oscillation (BW) will occur under proper conditions \cite{PhysRevA.81.033829}. The superposition of BO and ZT leads to the Bloch-Zener oscillation (BZO), and the superposition of BW and ZT leads to the breathing-Zener-wave-like oscillation (BZW). The properties of BZO and BZW in OWL will be investigated through the field-evolution analysis \cite{PhysRevA.81.033829, JOptSocAmB.27.1299.2010}. We inject a Gaussian beam \cite{PhysRevA.81.033829}
\begin{equation}
\psi(n,0)=(2\pi\sigma^2)^{-1/4} \exp\left [-\frac{(n - n_0)^2}{4 \sigma^2}-\mathbf{i}k_0(n - n_0)\right ]\,,
\end{equation}
with transverse wavenumber $k_0$ and intensity profile $|\psi(n,0)|^2$, which has a discrete Gaussian distribution centered at the $n_0$th waveguide with a spatial width $\sigma$. The exponential factor $\exp{[-\mathbf{i}k_0(n - n_0)]}$ captures the phase differences between the input beams excited at the $n$th and the $n_0$th waveguides. In what follows, we excite the array with a plane wavefront with $k_0 = 0$. We expand the input Gaussian beam in terms of the supermodes $|m\rangle$ as $|\psi (n, 0)\rangle = \sum_{m} A_m |m\rangle$, where $A_m = \langle m|\psi(n,0)\rangle$ is the expansion amplitude of the input beam. The subsequent wavefunction at a propagation distance $z$ is $|\psi (n, z)\rangle = \sum_{m} A_m e^{\mathbf{i}\beta_m z}|m\rangle$. The wavefunction in the reciprocal space can be obtained by taking the Fourier transform $|\phi (k,z)\rangle =\mathcal {F}[|\psi (n,z)\rangle ]$. The evolution of the beam intensity $|\psi(n, z)|^2$ in the real space and $|\phi(k, z)|^2$ in the reciprocal space along the propagation distance $z$ are used to illustrate the occurrence of BZO and BZW.


To observe BZO, we use a broad Gaussian beam with spatial width $\sigma = 3$ to excite the OWL from the central waveguide of the lower array. The parameters are $\alpha_u = \alpha_l = \alpha = 0.08$, $\delta_\beta = 0$ and $k_0 = 0$. Figure \ref{fig:GbothINlowWidth3}(a) shows the overlapped contour plots of $|\psi(n, z)|^2$ in the real space for the upper and lower arrays of PNC case, both of which are labeled from $1$ to $100$. Since the coupling constants $\kappa_1$ and $\kappa_2$ in the upper and lower waveguide arrays are positive and negative, respectively, the propagation of light beams in the two OWAs is opposite in the transverse direction \cite{OptLett.31.1322.2006}. From Fig.~\ref{fig:GbothINlowWidth3}(a), the light beam is reconstructed after two BO periods, thus the period of BZO $Z_{\rm BZO}$ and that of BO $Z_{\rm BO}$ satisfy the relationship $Z_{\rm BZO} = 2 Z_{\rm BO}$. In the reciprocal space (figures not shown here), ZT occurs at $k = 0.5\pi$ and $k=-0.5\pi$, which correspond to the spatial positions around $z_{T1} = Z_{\rm BO}/4$, $z_{T2} = 3Z_{\rm BO}/4$, $z_{T3} = 5Z_{\rm BO}/4$, \dots\,. When $z < z_{T1}$, the light beam mainly undergoes BO in the lower array. At $z = z_{T1}$, ZT occurs, whose tunneling rate is dependent on the width of the band gap. The light beam splits into two parts: one tunnels to the upper array, and the other one remains in the lower array. Due to the opposite transverse propagation directions in two arrays, the two BO paths are different and refocus at the second tunneling point $z = z_{T2}$. Since both bands are occupied, light beams tunnel between each other and coherently interfere. The net result is that more field tunnels to the upper array. At $z = z_{T3}$, more field tunnels back to the lower array. After the propagation distance $Z_{\rm ZBO}$, the light beam recovers. Due to the occurrence of anomalous BZO, the light beam tunnels at the mid zone between the upper and lower arrays periodically.

It has been proved theoretically \cite{OptLett.31.3351.2006} and experimentally \cite{PhysRevLett.96.023901.2006} that light evolution in the photonic lattice can reflect the band structure directly, whereas the width of each gap corresponds to the tunneling rate of ZT \cite{OptLett.31.3351.2006}. To demonstrate this principle clearly, we take half period ($Z_{\rm BO}/2$) of the light evolution in the real space and rescale by $N_h/\alpha$ for the waveguide index and $Z_{\rm BO}/(2\pi)$ for the propagation distance, as shown in Fig.~\ref{fig:GbothINlowWidth3}(b). The rescaled light evolution matches very well with the band structure described by Eq.~(\ref{eq:DISP}) shown by the solid line in Fig.~\ref{fig:OWLband}(b) and Fig.~\ref{fig:GbothINlowWidth3}(b). The BO path with larger oscillation amplitude occurs in the upper array, since its coupling constant is stronger which leads to a wider band. ZT is demonstrated by the exchange of field intensities of different BO paths at band hybridization, where the minimum gap opens.

For the PPC case, the evolution of broad Gaussian beam in OWL and its visual band are shown in Figs.~\ref{fig:GbothINlowWidth3}(c) and \ref{fig:GbothINlowWidth3}(d), respectively. From the band structure of PPC case [solid lines in Fig.~\ref{fig:GbothINlowWidth3}(d)], the two minibands hybridize around $k = 0.5\pi$ when $\delta_\beta=0$, but there is no ``true band gap'' throughout the whole Brillouin zone. Light beam can tunnel between the upper and lower arrays around $k = \pm 0.5\pi$, as shown in Fig.~\ref{fig:GbothINlowWidth3}(c). It indicates that BZO occurs in such kind of OWL with pure positive couplings. The different features between the light evolution of PPC case [Fig.~\ref{fig:GbothINlowWidth3}(c)] and PNC case [Fig.~\ref{fig:GbothINlowWidth3}(a)] are as follows. (i) The transverse propagation directions in the two layers of PPC case are the same since their couplings are both positive. (ii) The relation between periods of BO and BZO of PPC case is $Z_{\rm BZO} = 3 Z_{\rm BO}$. From the visual band for PPC case as shown in Fig.~\ref{fig:GbothINlowWidth3}(d), we can see that although there is no ``true band gap'', BZO still takes place with tunneling at the wavenumber position where the difference between the two minbands at the same $k$ reaches its minimum. This phenomena is attributed to the coherent properties of light waves, which is an advantage of optical system over other kinds of systems (electronic, plasmonic, elastic, etc).

For a single waveguide excitation ($\sigma = 0.02$), as shown in Fig.~\ref{fig:GbothINlowWidth002}, we observe BZW. Figs.~\ref{fig:GbothINlowWidth002}(a) and \ref{fig:GbothINlowWidth002}(b) demonstrate the contour plots of $|\psi(n, z)|^2$ in the real space for the upper and lower arrays, respectively. The light beam enters from the central waveguide of the lower array, refocuses at $Z_{\rm BO}$ and recovers at $Z_{\rm BZO}=2Z_{\rm BO}$ in the lower array. ZT occurs at $z_{T1} = Z_{\rm BO}/4$, $z_{T2} = 3Z_{\rm BO}/4$, $z_{T3} = 5Z_{\rm BO}/4$, ... In the upper array, the light beam cannot focus to a single waveguide and can reach a wider range of width due to its stronger coupling constant. In comparison, the refocused spatial range of the breathing mode of PPC case is much wider in the upper array [Fig.~\ref{fig:GbothINlowWidth002}(c)] and narrower in the lower array [Fig.~\ref{fig:GbothINlowWidth002}(d)].

The tunabilites of BZO (BZW) are enhanced by the band gap due to tunable hybridization in OWL. Comparing with the band gap in optical binary superlattices \cite{PhysRevLett.102.076802.2009}, which can be tuned in width vertically, the band gap in OWL can also be tuned in wavenumber position horizontally. Theoretically, we can tune the tunneling position from one array to the other one by varying the propagation constant difference $\delta \beta$ between two arrays and the tunneling rate by the strength of inter-array coupling constant $\kappa$. The spatial range and intensity of field evolution in OWL can also be tuned as well. Moreover, different from the fixed evolution pattern in a single OWA, the light beams have different evolution patterns in two arrays in OWL and can vary according to different tunneling positions. For the experimental verification, we can resort to the positive/positive coupling case, where the tunable hybridization still works and which is much easier to realize through the dielectric OWL.

In conclusion, we proposed a certain coupled optical waveguide
ladder and studied its band structure and Bloch-Zener oscillation. Due to the band hybridization through inter-array couplings, a band gap opens at the mid zone. This is completely different from the conventional Bragg gap. Thus, when a graded potential is applied, BZO (BZW) occurs with Zener tunneling at the mid zone.  It was found that the tunneling position and rate, spatial range and intensity of BZO (BZW) are tunable.
These research results are fundamental and important in understanding the coherent phenomena in lattice structures. This work will arouse interests from two highly active research communities in the fields of lattice systems and coherent effects.

\emph{Acknowledgments} This work was supported by RGC General Research Fund of the Hong Kong SAR Government. We acknowledge useful comments from Prof. K. Yakubo.



\clearpage

\section*{List of Figure Captions}

\noindent Fig. 1. (Color online) (a) Schematic diagram of a cross section of an optical waveguide ladder. The coupling constants for the positive/negative coupling case are $\kappa_1 = 1$, $\kappa_2 = -0.5$, and $\kappa = 0.2$. Band structures for various propagation constant differences (b) $\delta_\beta = 0$, (c) $\delta_\beta = 1$, and (d) $\delta_\beta = -1$. Band hybridization occurs at $k = 0.5\pi$, $0.4\pi$, and $0.6\pi$, respectively. The analytic results (solid lines) match well with the numerical results (discrete dots). (e) Wavenumber position of band hybridization as a function of propagation constant differences $\delta_\beta$ between two arrays. Fig1.eps

\begin{figure}[htbp]
  \centering
  \includegraphics[width=0.8 \textwidth]{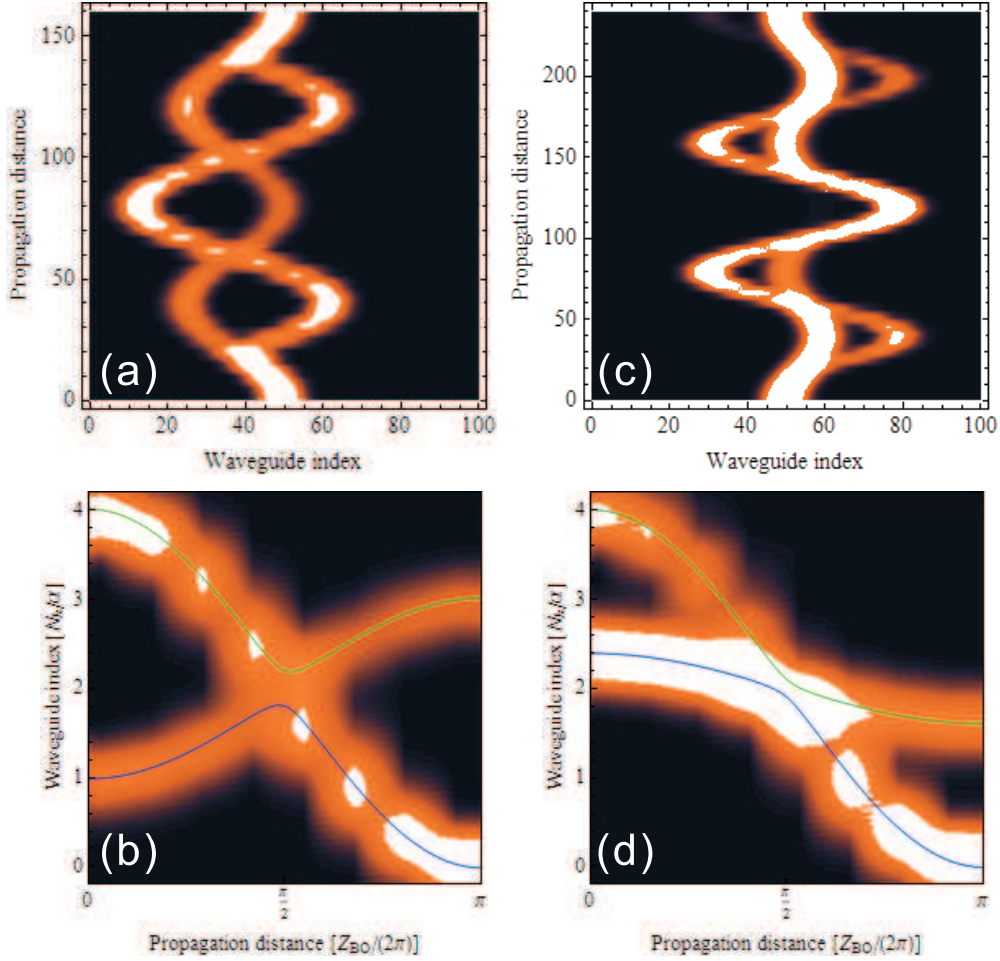}
  \caption{(Color online) Evolution of a broad Gaussian beam centered at $n_{\rm low} = 50$ with spatial width $\sigma = 3$. Positive/negative coupling case ($\kappa_1 = 1.0$, $\kappa_2 = -0.5$, and $\kappa = 0.2$): (a) Overlap of contour plots of $|\psi(n, z)|^2$ in the $n$-$z$ domain in the upper and the lower graded waveguide array. (b) Visual demonstration of band structure through spatial evolution of light beam. Positive/positive coupling case ($\kappa_1 = 1.0$, $\kappa_2 = 0.2$, and $\kappa = 0.1$): (c) is similar to Fig.~\ref{fig:GbothINlowWidth3}(a). (d) is similar to Fig.~\ref{fig:GbothINlowWidth3}(b). The common parameters are $N_{\rm up} = N_{\rm low} = 100$, $\sigma = 3$, $\beta_0 = 2$, $\delta_\beta = 0$, $\alpha = 0.08$.}
  \label{fig:GbothINlowWidth3}
  \end{figure}

\noindent Fig. 2. (Color online) Evolution of a broad Gaussian beam centered at $n_{\rm low} = 50$ with spatial width $\sigma = 3$. Positive/negative coupling case ($\kappa_1 = 1.0$, $\kappa_2 = -0.5$, and $\kappa = 0.2$): (a) Overlap of contour plots of $|\psi(n, z)|^2$ in the $n$-$z$ domain in the upper and the lower graded waveguide array. (b) Visual demonstration of band structure through spatial evolution of light beam. Positive/positive coupling case ($\kappa_1 = 1.0$, $\kappa_2 = 0.2$, and $\kappa = 0.1$): (c) is similar to Fig.~\ref{fig:GbothINlowWidth3}(a). (d) is similar to Fig.~\ref{fig:GbothINlowWidth3}(b). The common parameters are $N_{\rm up} = N_{\rm low} = 100$, $\sigma = 3$, $\beta_0 = 2$, $\delta_\beta = 0$, $\alpha = 0.08$. Fig2.eps

\begin{figure}[htbp]
  \centering
  \includegraphics[width=0.8 \textwidth]{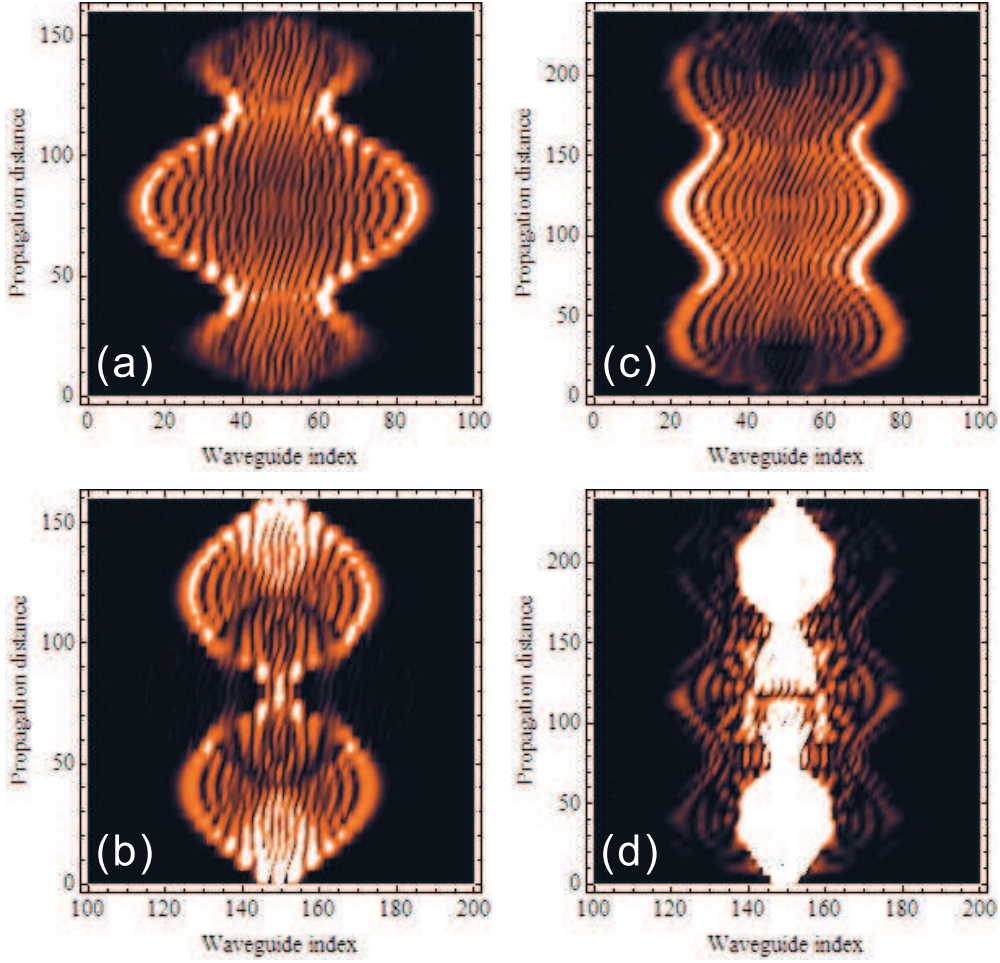}
  \caption{(Color online) Evolution of single waveguide excitation starting from the lower array center $n_{\rm low} = 50$. Positive/negative coupling case ($\kappa_1 = 1.0$, $\kappa_2 = -0.5$, and $\kappa = 0.2$): (a) Contour plots of $|\psi(n, z)|^2$ in the $n$-$z$ domain in the upper graded waveguide array and (b) that in the lower graded waveguide array, respectively. Positive/positive coupling case ($\kappa_1 = 1.0$, $\kappa_2 = 0.2$, and $\kappa = 0.1$): (c) is similar to Fig.~\ref{fig:GbothINlowWidth002}(a), (d) is similar to Fig.~\ref{fig:GbothINlowWidth002}(b). Parameters are the same as those in Fig.~\ref{fig:GbothINlowWidth3} except for $\sigma = 0.02$.}
  \label{fig:GbothINlowWidth002}
  \end{figure}

\noindent Fig. 3. (Color online) Evolution of single waveguide excitation starting from the lower array center $n_{\rm low} = 50$. Positive/negative coupling case ($\kappa_1 = 1.0$, $\kappa_2 = -0.5$, and $\kappa = 0.2$): (a) Contour plots of $|\psi(n, z)|^2$ in the $n$-$z$ domain in the upper graded waveguide array and (b) that in the lower graded waveguide array, respectively. Positive/positive coupling case ($\kappa_1 = 1.0$, $\kappa_2 = 0.2$, and $\kappa = 0.1$): (c) is similar to Fig.~\ref{fig:GbothINlowWidth002}(a), (d) is similar to Fig.~\ref{fig:GbothINlowWidth002}(b). Parameters are the same as those in Fig.~\ref{fig:GbothINlowWidth3} except for $\sigma = 0.02$. Fig3.eps


\clearpage


\newpage
\singlespacing

\newpage

\newpage

\end{document}